\documentclass[aps,prd,twocolumn,groupedaddress,showpacs,preprintnumbers,nofootinbib,draft]{revtex4}
\usepackage{epsf}
\usepackage{bm}
\usepackage{amsfonts}
\usepackage{colordvi}
\usepackage{graphicx}
\usepackage{latexsym}

\newcommand{\be}{\begin{equation}}
\newcommand{\ee}{\end{equation}}
\newcommand{\beqy}{\begin{eqnarray}}
\newcommand{\eeqy}{\end{eqnarray}}
\newcommand{\p}{\partial}

\newcommand{\mx}{\mbox}

\newcommand{\bb}{\beta}

\newcommand{\e}{\epsilon}

\newcommand{\la}{\lambda}
\newcommand{\La}{\Lambda}

\newcommand{\ra}{\rightarrow}

\newcommand{\im}{\Longleftrightarrow}

\newcommand{\vs}{\vspace{5mm}\\}
\def\be{\begin{equation}}
\def\ee{\end{equation}}
\def\ba{\begin{eqnarray}}
\def\ea{\end{eqnarray}}
\begin{document}
\title{Can Inflation solve the Hierarchy Problem?}
\author{Tirthabir Biswas\footnote{tirtho@physics.mcgill.ca}   and Alessio Notari\footnote{notari@hep.physics.mcgill.ca}}
\affiliation{Physics Department, McGill University, 3600 University Road, Montr\'eal, QC, H3A 2T8, Canada}

\date{\today}

\begin{abstract}
Inflation with
tunneling from a false  to a true vacuum becomes viable in the presence
of a scalar field  that slows down the initial de Sitter phase.
As a by-product this field also sets dynamically the value of
$M_{Planck}$ observed today. This can be very large if the tunneling
rate (which is exponentially sensitive to the barrier) is small enough. Therefore along with Inflation we also provide a natural dynamical explanation for why gravity is so weak today. Moreover we predict a spectrum of gravity waves peaked at around 0.1 mHz, that will be detectable by the planned space inteferometer LISA.
 Finally we discuss interesting predictions on cosmological scalar and tensor fluctuations in the light the WMAP 3-year data.
\end{abstract}


\maketitle

\section{Introduction}

At least two fundamental scales appear to be present in the observed Universe, and they are extremely different: the electroweak scale, $M_{EW}$, and the Planck scale $M_{Pl}$.
The fact that their ratio appears to be around $M_{EW}/M_{Pl}\approx 10^{-14}-10^{-15}$ is a puzzle for many reasons.

First, one can have the theoretical prejudice that a deeper comprehension of physics should lead us to a theory with one single mass scale. So the fact that gravity is so much weaker than other forces of Nature seems a problem whose resolution will lead us to a better understanding of our Universe.

Second, even if we assume that the fundamental theory has two different mass scales, one has to understand what is there  in the ``desert''  between these two scales, and at which scale new physics will appear? This is a very important question both for experimental purposes (is it worth building accelerators to explore this desert?) and for theoretical problems.
In fact, the new physics scale is assumed to set the ultraviolet cutoff for the presently known particle physics.
It is well known that the Standard Model of particle physics suffers from a major theoretical problem, which is the stability of the Higgs mass under radiative corrections: the Higgs mass is quadratically sensitive to the ultraviolet cutoff and if the cutoff scale is much higher than the electroweak scale an extreme fine-tuning between the bare mass and the one-loop correction is required to give a low value for the physical mass (we know from electroweak precision data that the latter cannot exceed about 200 GeV, for a review see \cite{lep} and references therein).
It is more than plausible therefore that the new physics scale has to be very close to $M_{EW}$.
However the problem could still persist going up to the Planck scale, which is the highest known scale, unless the new physics is able to ``screen'' the sensitivity to $M_{Pl}$. This possibility is the main motivation for models of low-scale supersymmetry. However no hint for this possibility has been found in accelerators until now, and the arrival of the LHC cries for other possibilities.

An alternative possibility, which has become very popular in the recent years, starts with the idea that there may only be a {\it single} scale in the fundamental theory. If this scale is $M_{EW}$ (rather than $M_{pl}$), then the cutoff scale would be very close to the experimental bound (which is around $10$ TeV, \cite{lep})\footnote{Note that the main point of this paper is the hierarchy between gravity and electroweak scale, and we do not address the problem of the 'Little'' Hierarchy of about two orders of magnitude between this cutoff scale and the Higgs mass. }.

This has been proposed in the framework of extra dimensional models \cite{ADD,RS}: in such scenarios there is only one fundamental scale and the weakness of gravity comes from the fact that only gravity propagates in the ''bulk''.

The new proposal that we put forward here is similar in spirit but more conservative. We also assume that there is only {\it one} fundamental scale, and (similarly to  \cite{RS}) we exploit the presence of an exponentially sensitive quantity in order to explain the Hierarchy Problem. However we need no extra dimensions and moreover we get the explanation as a by-product of Inflation, which is another major paradigm \cite{Guth} that requires physics beyond the standard model, making this scenario even more interesting.

A new idea for Inflation has been put forward in~\cite{notaridimarco}, where it was shown that a false vacuum can give rise to a viable model of Inflation, with the addition of one scalar field with  non minimal coupling to gravity. Such a field slows down dramatically the exponential expansion of the Universe, allowing graceful exit through tunneling and bubble nucleation when the Hubble parameter becomes of the order of the tunneling rate. In this way the fact that the Universe could have reasonably started from a false vacuum  turns out to be what gives our post-inflationary Universe.
This realizes inflation without flat potentials, which are the framework for slow-roll inflation.

Now, our proposal for a solution of the Hierarchy problem arises from this idea, and it is radically different from supersymmetry or extra dimensions. It is a dynamical solution, which makes the Hierarchy a {\it time dependent} quantity, that evolves during Inflation.

The small number $M_{EW}/M_{Pl}$ is explained via another small number $\Gamma_{vac}^{1/4}/M$, where $M$ is the only fundamental scale (around  TeV), and $\Gamma_{vac}$ is the tunneling rate per unit volume of the false vacuum to the true vacuum, which is assumed to be at almost zero energy. The crucial point is that $\Gamma_{vac}$ is exponentially sensitive to the details of the potential \cite{Coleman}, and therefore it can easily be many orders of magnitude smaller than the fundamental scale $M$.

Remarkably our idea is clearly testable, having a number of predictions, observable in the near future. There are in fact very few relevant parameters and the model is already very constrained.

The most striking prediction is a distinctive spectrum coming from bubble collisions at reheating. In our model the reheating temperature is close to the TeV scale and this determines the frequency of the peak of gravity waves. It happens that it falls in the range of sensitivity of the space interferometer LISA \cite{lisa}, with a detectable amplitude. 
Moreover the frequency corresponds directly to the energy scale of new physics, and the latter could be eventually measured also by the LHC, as the cutoff scale of higher dimensional operators. This would allow us to compare two completely different observations, regulated by the same scale, therefore providing an extremely clear test to confirm or rule out our idea.

In addition to this, we discuss also the power spectrum of tensor (gravity waves) and scalar (density) perturbations, that any inflationary model produces. Depending on the specific realization of our idea we can make a prediction either on the scalar spectral index\footnote{In the first version of this paper we made a prediction on the scalar spectral index which is in good agreement with the new WMAP 3-year data~\cite{WMAP3}.}, or on the 
stochastic background of gravity waves produced by quantum fluctuations during inflation measurable by up-coming experiments.

To summarize, the addition of a single scalar field to ``Old Inflation'' achieves  the following: \\
\noindent
$\bullet\,$ it  provides a graceful exit to Inflation, \\
\noindent
$\bullet\,$ it does not need to assume flat potentials, and it explains easily the horizon and flatness problems together with giving a spectrum of cosmological perturbations in agreement with WMAP, \\
\noindent
$\bullet\,$ it explains dynamically the Hierarchy problem, using the fact that a tunneling rate is naturally exponentially suppressed, \\
\noindent
$\bullet\,$ it gives rise to a model of inflation which is viable and testable in the near future (especially through gravity waves).

\section{{\bf  Inflation and Hierarchy}}
The model that we are going to study has an unstable vacuum energy, that we will call $\Lambda$ with some tunneling rate per unit volume $\Gamma_{vac}$. The only other ingredient in our model is a scalar field $\phi$ which has a generic coupling to the Ricci scalar and a potential $U(\phi)$.

So, the action is:
\begin{eqnarray}
 S &=& \int d^4 x \sqrt{-g} 
\left[ \frac{1}{2}  M^2 f(\phi) R - \frac{1}{2} \partial_{\mu}\phi \partial^{\mu}\phi  \right. \nonumber \\
 && \left.
 - U(\phi) - \Lambda -{\cal L}_{SM} \right] \label{lagrangiana} \, ,
\end{eqnarray}
where we consider $f(\phi)>0$ and we assume that there is only one fundamental scale (close to $10$ TeV) appearing in  $U(\phi)$, $f(\phi)$ and $\Lambda$. We parameterize everything in terms of $M$, and then we explain later (sect.\ref{fifth}) how close $M$ can be to $10$ TeV. For example we parameterize $\Lambda \equiv \lambda^4 M^4$. As we will see, however, we require some tuning of order $10^{-3}$ on $\lambda$ (in this paper by ``tuning'' we will always refer  to hierarchy between mass scales) in order to produce the correct amplitude for cosmological perturbations. 

For our mechanism to work we essentially require that $f(\phi)$ has a minimum, say $f(0)=1$, and then increases monotonically faster than $\phi^2$, as $\phi \to \pm\infty$. Simple  examples that one can  keep in mind  are $f=1+\bb(\phi/M)^n$, with $n>2$, and $f=\cosh(\bb\phi/M)$. Finally ${\cal L}_{SM}$ is the usual Standard Model lagrangian of particle physics, whose cutoff scale is assumed to be around $10$ TeV, and which does not play any dynamical role in our scenario.

Qualitatively what happens in this model is quite simple. We assume the field $\phi$ to start close to zero (we explain how such initial conditions can be realized naturally in section \ref{exponential}). We ignore the potential $U(\phi)$ for the moment: we will use it only for stabilizing the field at late times after reheating, and we assume it is subdominant with respect to $\Lambda$. Therefore what happens is that the presence of a vacuum energy drives exponential inflation with Hubble constant $H_I$. At the same time the field $\phi$ is unstable and starts growing in such a background due to its coupling to $R$, which is large during inflation. When $\phi$ becomes comparable to $M$, the de Sitter phase is dramatically slowed down and in fact we reach an asymptotic regime, in which $\phi$ grows to arbitrarily large values and the expansion is power law \cite{dolgovford,notaridimarco}. (We clarify in section \ref{asint} why in this phase, although $\phi\gg M$, quantum gravitational corrections are not expected to spoil the picture.) This means that the Hubble constant $H$ is decreasing during this period and when it reaches the value $\Gamma_{vac}^{1/4}$, the false vacuum decays rapidly nucleating bubbles of true vacuum and giving rise to a thermal bath (through bubble collisions) and thus to the usual radiation era.  We do not have to worry about early production of bubbles here (that could spoil the CMB isotropy), since the $\Gamma/H_I^4$ that we require here is extremely small and by many orders of magnitude below the experimental bounds (see \cite{notaridimarco}).

This  mechanism was already described in~\cite{notaridimarco} for the special case when $f(\phi)$ is quadratic.  Here we point out how  the model discussed in \cite{notaridimarco} can be naturally  generalized to give us an interesting by-product: one realizes that the bare mass $M$ in the Lagrangian does not coincide with the value of $M_{Pl}$ that we observe today, but the latter is determined in general by the equation:
\be
M_{Pl}^2=M^2 f(\phi)>M^2 \left({\phi\over M}\right)^2 \, ,\ \phi\gg M \label{gerarchia}  \, ,
\ee
which follows from the assumption we have made on $f(\phi)$. The dynamical evolution that we described can drive $f(\phi)$ to extremely big values. The final value of $f(\phi)$ is determined by the condition $H=\Gamma_{vac}^{1/4}$. Therefore if $\Gamma_{vac}$ is small enough, $f(\phi)$ has  enough time to grow to ensure $M_{Pl}\gg M$.
So, the basic idea is the following. We assume $M$ to be the TeV scale, which is the fundamental scale also for ${\cal L}_{SM}$. This means that gravity at the beginning of Inflation was much stronger than today.
Then, the large hierarchy between $M_{Pl}$ and $M$ is generated by the evolution of the $\phi$ field during inflation and its final value is set by another small parameter $\Gamma_{vac}^{1/4}/M$.

The fact that $M_{Pl}$ becomes big after inflation and all other scales present in the matter Lagrangian ${\cal L}_{SM}$ stay at the value $M$ is very clear in the action of eq.(\ref{lagrangiana}). However at this point we find it easier to go from the so-called Jordan frame of eq.(\ref{lagrangiana}) to the  Einstein frame, in which all the fields are rescaled in such a way that gravity appears in the usual Einstein form and the field $\phi$ couples to everything else.

We also said that we have only one fundamental scale and  this is around TeV, but actually the second part of the statement is frame dependent: in the Einstein frame the fundamental scale $M$ is identified with the Planck scale, and what happens during inflation is that the matter Lagrangian gets suppressed by a huge factor, so that at late times particle physics apppears to be at a very small scale, compared to $M$. It does not really matter in which frame we are working since the physical observables are always ratios of two scales, and this does not depend on the frame. So, the frame independent statement is that in our model the ratio $M_{EW}/M_{Pl}$ is driven to a very tiny value by the dynamics of the Universe in the false vacuum.
\section{Evolution in the Einstein frame}
The action in eq.(\ref{lagrangiana}) is dynamically equivalent to a theory in which the gravitational action is the usual one, via the conformal transformation:
\be
\bar{g}_{\mu\nu}=f(\phi)g_{\mu\nu}\, ,
\ee
where we use the bar to indicate a quantity in the new frame.
The new action looks like
\be
S_{E}= {1\over 2}\int d^4x\ \sqrt{-\bar{g}}[M^2 \bar{R}-K(\phi)(\bar{\p}\phi)^2]
\label{e-action} \, ,
\ee
where,
\be
K(\phi)\equiv {2f(\phi)+3  M^2 f^{'2}(\phi)\over 2f^2(\phi)}
\label{K} \, .
\ee
We have neglected the potential $U(\phi)$, since it is assumed to be subdominant relative to the false vacuum energy, which   after the conformal transformation  becomes
\be
-S_{vac}=\int d^4x\ \sqrt{-\bar{g}}\frac{\Lambda}{f^{2}(\phi)} \equiv \int d^4x \sqrt{\bar{g}}\, \bar{V}(\phi)\, .
\ee
So, in this frame the false vacuum energy becomes a potential term for $\phi$, that we called $\bar{V}$.
\subsection{Exponential Phase}\label{exponential}
In order to understand the dynamics, we study two different limits separately. First consider the case when $\phi \ll M$. In this limit, generically the power series expansion of $f(\phi)$ can be approximated as
\be
f(\phi)\approx 1 + \beta\left( \frac{\phi}{M} \right)^n \, .
\ee
Since we have assumed that $f>0$ has a minimum at $\phi=0$, $n\neq 1$ and $\bb>0$.
The renormalizable coupling ($n=2$) is expected to be the dominant one and so we study this case here. However everything works also if the $n=2$ term is absent, and we discuss this case in Appendix \ref{n>2}.

 Now,
 keeping the lowest order terms in the kinetic and the potential terms amounts to having
\be
K(\phi)=1+\bb(6 \beta-1)\left(\frac{\phi}{M} \right)^2 \, , \, \bar{V}\approx \Lambda \left[1-2\bb\left(\frac{\phi}{M} \right)^2 \right] \, .
\ee
One can now introduce a canonical scalar field variable $\Phi$ via
\be
\sqrt{K(\phi)}d\phi =d\Phi
\label{Phi-def} \, ,
\ee
so that the final Einstein frame action looks like
\be
S= \int d^4x\ \sqrt{-g}\left[{M^2\over 2}R-{1\over 2}(\p\Phi)^2-\bar{V}(\Phi)\right]
\label{canonical} \, ,
\ee
where $\phi=\phi(\Phi)$ is given by (\ref{Phi-def}). 

The vacuum term $\Lambda$ becomes in this frame a potential $\bar{V}$ that looks like the top of a hill, and  therefore we will get inflation if we start from close to the maximum ($\phi\ll M$): so, in this frame things  
look like slow-roll inflation.

This may seem like hybrid inflation(\cite{hybrid}), with inflation ending through tunneling.
However there are two differences. First, hybrid inflation is introduced assuming directly some specific couplings between two fields: after rolling in one field, the barrier along the other direction is lowered, giving graceful exit. Here instead in the original frame of eq.(\ref{lagrangiana}), the $\phi$ field and the field locked in the minimum are not coupled: they happen to be coupled in the Einstein frame because $\phi$ is universally coupled to matter.
Second, here the slow-roll phase is followed by a power-law phase: so even if inflation in this frame happens at high scale ($10^{-3} M_{Pl}$), the reheating scale is very low (TeV): the field tunnels because $H$ decreases, not because the barrier lowers.

It is easy to see also how the initial condition on $\phi$ can naturally arise. 
One possibility is that the initial condition is just $\phi$ close to zero everywhere. Another possibility (as was first argued in \cite{topological}) is that as initial condition $\phi$ takes random values in different regions of space. Similar to the case of topological inflation, in this case one expects domain wall like configurations to exist extrapolating between $\phi\ra \pm\infty$. Thus there would be regions where $\phi\approx 0$  which are going to inflate and eventually overwhelm the universe (Regions which start with larger values of $\phi$  are  going to inflate much less, if at all.).

Since this is going to be slow roll inflation (in this frame), one can use the slow roll approximations. One can check that in the slow roll approximations, to lowest order in $\phi/M$, $K(\phi)$ can be approximated by $1$, and the approximate potential for $\Phi$ is given by
\be
\bar{V}(\Phi)=\Lambda\left[1-2\bb\left({\Phi\over M}\right)^2\right] \, .
\ee
It is now an easy exercise to calculate the slow roll parameters:
\be
\e\equiv{M^2\over 2}\left|{1\over V}{dV\over d\Phi}\right|^2=8 \bb^2 \left(\frac{\Phi}{M} \right)^{2} \, ,
\ee
\be
\eta\equiv M^2{1\over V}{d^2V\over d\Phi^2}=-4\bb \label{epseta} \, .
\ee
In the slow-roll approximation the relation between the number of efolds ${\bar{\cal N}}$ (defined to be zero when $\beta \phi^2 \approx M^2$) and $\phi$ is given by:
\be
\bar{{\cal N}}\approx \frac{1}{4 \beta}\ln\left[\frac{M}{\sqrt{\beta}\phi}\right] \, .
\label{enne}
\ee 
When $\phi$ grows to become of the order of $M$ the exponential phase ends, and there is a transition to a power-law phase.

At this point it is perhaps good to emphasize that although the inflation itself is driven by the energy of the false vacuum (of possibly another  scalar field), the fluctuations are generated in the $\phi$ field. 
\subsection{{\bf Power Law Phase}} \label{asint}
When $\phi$ grows to become larger than $M$, all possible nonrenormalizable operators which might be there in the theory become relevant. We exploit this fact in order to slow down strongly\footnote{If one only had the quadratic coupling even at large $\phi$, it  still gives a good model of inflation \cite{notaridimarco}, but in the power law phase $\phi$ does not grow fast enough to drive the Planck mass to very large values. Moreover the presence of the nonrenormalizable operators makes unnecessary to have a curvaton, which was required in \cite{notaridimarco} in order to have a flat adiabatic spectrum.} the inflationary phase, and we only make the assumption that 
$$f(\phi)>\phi^2\im f^{'2}>|f| \, .$$ 
In this case in the numerator of $K(\phi)$ (\ref{K}), the second term dominates.
Therefore, in this phase, the kinetic and potential terms can be approximated as
\be 
K(\phi)\approx {3\over 2}\left(f'\over f\right)^{2}
\, , \qquad
\bar{V}\approx {\La\over f^{2}} \, .
\ee
We can now introduce a canonical variable via
\be
\Phi \equiv  \sqrt{3\over 2} M\ln f \, ,
\ee
so that the scalar tensor action  looks formally the same as eq.(\ref{canonical}) with
\be
\bar{V}(\Phi)=\Lambda\exp\left({-2\sqrt{\frac{2}{3}}{\Phi\over M }}\right) \, .
\label{potential}
\ee
Evolutions under such exponential potentials are very well known \cite{copeland} and in particular, this corresponds to a power law phase given by
\be
\bar{a}\sim \bar{t}^p \mx{ with }p={3\over 4} \, .
\ee
It is also easy to see that $\phi$ grows and the kinetic energy is always proportional to $\bar{V}$ (precisely it is $4/5 \, \bar{V}$).

Now, the end of this phase is achieved when $\bar{H}^2\simeq \frac{\bar{V}}{M^2}$ is equal to $\bar{\Gamma}_{VAC}^{1/2}$. Therefore the final field value $\phi_F$ is given by:
\be
f(\phi_F)\simeq \frac{\lambda^2 M }{\bar{\Gamma}_{VAC}^{1/4}}  \label{phif} \, .
\ee

At this point some readers may have the objection that a theory in which the field value is much bigger than the fundamental scale is suspicious, since quantum gravitational effects may invalidate our treatment. However this not a new situation in Inflationary cosmology and in fact the so-called ``chaotic inflation'' \cite{chaotic} has always the inflaton value bigger than $M_{Pl}$. Such a large value for the field is not a problem and one can use classical gravity since we are always working at sub-Planckian energies. In other words, quantum corrections to the Lagrangian always involve a potential $V(\Phi)$ or a mass $\partial^2 V(\Phi)/\partial \Phi^2$ and never directly $\Phi$ (see \cite{LindeBook}, sect.2.4 for a clear discussion).  Since $\bar{V}$ is always smaller than $M$, our model is well under control, and actually the situation becomes better and better as $\phi$ increases.
\subsection{Hierarchy, Inflationary constraints and Gravity Waves}  \label{constr}
First of all, let us see how much Hierarchy is generated. This can be seen from eq.(\ref{gerarchia}), where we have to put the final value given by eq.(\ref{phif}). This gives:
\be
\frac{M_{EW}}{M_{Pl}} \simeq \frac{1}{\lambda} \sqrt{\frac{\bar{\Gamma}_{VAC}^{1/4}}{M}} \, .
\ee
Note now that $\bar{\Gamma}$ is different from $\Gamma$ in the original frame. In fact $\bar{\Gamma}$ is a time dependent quantity as any other mass scale in the Einstein frame. Therefore if we want to relate the Hierarchy to the time independent $\Gamma$ in the original frame we have to take into account the fact that any mass scale $m$ is suppressed in the Einstein frame as:
\be
\bar{m}= f^{-1/2}(\phi_F) m \, .
\ee
Therefore we get the following which is the main result of this paper:
\be
\frac{M_{EW}}{M_{Pl}} \simeq \frac{1}{\lambda^2} \frac{\Gamma_{vac}^{1/4}}{M} \, ,
\ee
that relates the hierarchy to the smallness of the tunneling rate.

We are going to show, then, under which conditions our model is also satisfactory as a model of primordial inflation, in the light of the recent WMAP 3-year data~\cite{WMAP3}. This will set the value of the parameters $\beta$ and $\lambda$.

First of all, let us compute when a particular comoving scale $L$ went outside the horizon during inflation. We count the number of efolds starting from the end of exponential inflation (whose scale factor we call $a_E$), going backwards in time. In general a scale $L$ leaves the horizon at some efolding number $\bar{{\cal N}}_{L}$ if:
\be
L \left(\frac{T_0}{T_{\rm{RH}}} \right)\left(\frac{\bar{a}_{E}}{\bar{a}_{RH}} \right) e^{- \bar{{\cal N}}_L}=\bar{H}_{I}^{-1} \, ,
\ee
where $\bar{H}_I\simeq\lambda^2 M$ is the value of $\bar{H}$ during the exponential phase.
Here the reheating temperature $T_{RH}$ is the temperature of the thermal bath after the bubble nucleation. Assuming that reheating is almost istantaneous it is given by $T_{RH}^4\simeq \Gamma_{vac}^{1/2} M_{Pl}^2$. The redshift during the power-law phase is given by $\bar{a}_{E}/\bar{a}_{RH}=(\bar{t}_{E}/\bar{t}_{RH})^{3/4}\simeq(\bar{\Gamma}^{1/4}_{VAC}/\bar{H}_I)^{3/4}$. Here we have assumed that the transition between the ``exponential'' and the ``power-law'' phase is very quick and this is true for most functions, $f(\phi)$. Now, the largest scale observed today is the horizon scale ($3000 h^{-1} Mpc$), which leads to:
\be
\bar{{\cal N}}_{3000 h^{-1} Mpc}\approx 49+ \ln \lambda \, .  \label{orizzonte}
\ee
Thus it is clear that the horizon problem will be solved if the total number of e-foldings, $\bar{{\cal N}}_{tot}\gtrsim 46$ (where we used $\ln\lambda\simeq -3$, given by eq.(\ref{reasonable})).

Next, we proceed to analyze the spectral index of cosmological perturbations. This is given by $n_S=1+2\eta-6\epsilon$. Using eq.(\ref{epseta}) we get:
\be
n_S-1\simeq 2 \eta \simeq -8\beta  \, ,  \label{ns}
\ee
in agreement with the Jordan frame computation in~\cite{notaridimarco}.

Using the most recent observations by the 3-year data of the WMAP satellite\cite{WMAP3} we can fix the coupling $\beta$ to be:
\be
n_S\simeq 0.95 \Rightarrow \beta \simeq 6 \cdot 10^{-3}  \label{betaWMAP} \, 
\ee


The slow roll parameter $\epsilon$ is instead much smaller:
\be
\epsilon\simeq 8 \beta e^{-4 \beta \bar{{\cal N}} } \simeq 4.8 \times 10^{-2} e^{-0.024 \bar{{\cal N}}}  \label{epsn=2} \, ,
\ee
where we used eq.(\ref{enne}) for small $\phi$, and in the second equality eq.(\ref{betaWMAP}).
Given these values for the slow-roll parameters we can evaluate also the amplitude of the power spectrum , which is given by the well-known formula (see {\it e.g.} \cite{LindeBook}):
\be
A^2=\left. \left(\frac{\bar{H}_I}{M}\right)^2\frac{1}{8 \pi^2 \epsilon} \right\arrowvert_{\phi=\phi(\bar{{\cal N}}\approx \bar{{\cal N}}_{3000 h^{-1} Mpc})}  \label{ampiezza} \, .
\ee
If we insert eq.(\ref{epsn=2}) in this, combined with eqs.(\ref{betaWMAP},\ref{orizzonte}), and we impose it to be of the observed order $10^{-10}$ we get a constraint on $\lambda$
\be
\lambda \simeq 3\times 10^{-3}  \label{reasonable} \, .
\ee

Let us also check if the horizon problem is solved successfully.
 Now, the total number of efolds $\bar{{\cal N}}_{tot}$ depends on the initial condition of our field $\phi=\phi_0$.
 We assume as initial condition that $\phi$ starts close to zero. It is not physical to assume it to be exactly zero since the field is subject to quantum fluctuations. Therefore we set $\phi_0$ to be determined by initial quantum fluctuations. In this case one has $\phi_0\sim \bar{H}_I/\sqrt{\beta}$. This gives:
\be
\bar{{\cal N}}_{tot}\approx -\frac{\ln \lambda }{2 \beta}\simeq 484  \, .
\ee
which is more than enough.

The model is now very constrained and so we have a definite prediction on the amplitude of the spectrum of stochastic gravitational waves produced (as usual) by quantum fluctuations during inflation.
The tensor-to-scalar ratio can be computed in the Einstein frame just like the usual one for slow-roll inflation:
\be
r\equiv \left.\frac{P_{T}}{P_S}\right\arrowvert_{\bar{  {\cal N}}  =\bar{{\cal N}}_{3000 h^{-1} Mpc}}
=16 \epsilon\arrowvert_{\bar{{\cal N}}=\bar{{\cal N}}_{3000 h^{-1} Mpc}}  = 0.25 \, ,
\ee
which is just below the present constraints, and so it is going to be measured by coming experiments~\cite{Cooray}.

It is very interesting to note what happens in the case $n>2$, which is analyzed in detail in 
Appendix \ref{n>2}. There is no constraint on $\beta$ analoguous to eq.(\ref{betaWMAP}): in fact the spectral index is independent of $\beta$ and it is only weakly dependent on $n$. Remarkably the resulting $n_S$, given in eq.(\ref{prediction}) is a prediction ($n_S\simeq 0.95$) in very good agreemeent with the WMAP recent measurement (\cite{WMAP3}). This case
requires  $\la$ to be tuned approximately by the same amount as eq.(\ref{reasonable}). The duration of inflation is enormously long and $P_{T}/P_S$ is unobservable, since $\epsilon$ is extremely tiny in this case.

Finally, and very importantly, we stress that in our scenario the reheating temperature is quite low:
$T_{RH}\simeq\left(\Gamma_{VAC}^{1/2} M_{Pl}^2\right)\approx \sqrt{M M_{EW}} \lambda$. This is in fact close, or less than a TeV (we will be more precise in the next section), although the scale of inflation in this frame is very high ($10^{-3} M_{Pl}$). Note that this could mean that the electroweak symmetry is either restored for a very small time or never restored in the history of the Universe. 

This has a very exciting observable consequence. In fact, since reheating proceeds through bubble collisions, it is well-known \cite{GW} that relic gravitational waves (GW) are produced. For a recent analysis of the GW spectrum we refer to~\cite{GW}: 
a rough estimate gives that the spectrum of GW is peaked around $10^{-1}-10^{-3}$ mHz (which is in the sensitivity range of LISA), and the amplitude is big enough to be detectable by LISA (see \cite{GW,lisa}). Although a precise estimate is beyond the scope of this paper, one can use the equations in \cite{GW} and find that the amplitude of the peak should be around $\Omega_{GW}\simeq 10^{-7}$.

\section{{\bf Late time Stabilization and ``Other'' Hierarchies}} \label{fifth}

If after the decay of the false vacuum we are left with no potential for $\Phi$, it will mediate a fifth force among the SM particles grossly violating the current experimental bounds. There may be different mechanisms which may solve this problem but here we point out the simplest one, which is stabilizing and giving it a mass through the potential $U(\phi)$, introduced in eq.(\ref{lagrangiana}).  As a concrete example one may imagine a periodic potential (although the only necessary requirement is that the potential have many minima):
\be
U(\phi)=-\mu^4 M^4 \sin^2 \left[ f(\phi) \right] \, , \label{U}
\ee
where $\mu$ is a number that we will assume to be slightly smaller than $\lambda$, so that the potential does not change in a significant way the dynamics previously discussed. Using eq.(\ref{reasonable}) this means:
\be
\mu \lesssim 10^{-4} \, .  \label{Uconst}
\ee
In the Einstein frame the potential becomes
\be
\bar{U}(\phi)=-\mu^4 M^4 \frac{\sin^2 \left[f(\phi)\right]}{f^2(\phi)} \, .
\ee
We see that as long as $\mu$ is small compared to $\lambda$,
 this potential remains subdominant through out the evolution of $\phi$ and plays no significant role in the dynamics\footnote{We have explicitly checked that a field rolling in a potential with bumps of small amplitude gets only small corrections to its kinetic energy}.
However, once $\Lambda$ vanishes, $\phi$ stops and it falls in a minimum of the potential around the value $\phi_F$. This leads to a mass of the order:
\be
m_{\Phi}=\sqrt{\frac{\partial^2 \bar{U}(\Phi)}{\partial\Phi^2}}\sim {\mu^2 M_{Pl}}  \, ,
\ee
which easily avoids the fifth-force bounds. This is also big enough to avoid any problem of overclosure of the Universe in the radiation phase due to  oscillations in the  $\phi$ field, which scales as matter. The gravitational decay rate is big enough so that the $\phi$ oscillations decay away before nucleosynthesis \cite{moduli}.

At this point we can be more precise about the value of $M$ in the original frame, in order to discuss more in detail  the amount of tuning needed and the prediction for the LISA experiment.
We said that $M$ has to be close to 10 TeV, but one can imagine two specific situations. 
In the first one, $M=$10 TeV, which means that the scale of $\Lambda$ is about $(10 {\rm GeV})^4$. In this case the frequency of the gravity wave would be $0.1$ mHz, where the sensitivity of LISA is very good ($\Omega_{GW}\simeq 10^{-10}$).
This choice leads to the issue of finding a motivation for why the scale of $\Lambda$ and of $U(\phi)$, 
are kept smaller than the cutoff scale. This is beyond  the scope of this paper, but one can imagine~\cite{BN2} for example an axion-like potential where the scale is low because it is dynamically generated (as for the QCD axion).
In the second situation one can imagine that $\Lambda^{1/4}$ coincides with the  10 TeV cutoff scale. In this case there is no issue of protecting $\Lambda^{1/4}$ and $U(\phi)$ from quantum corrections, but the issue is instead: why $M$ is larger than the cutoff scale already at the beginning of inflation? In this case one could think of something similar to what happens in string theory, where the four dimensional  Planck scale can be easily 2-3 orders of magnitude bigger than the fundamental string scale. In this second case the prediction for LISA is that the spectrum is peaked around $10^{-3}$ mHz, with the same amplitude $\Omega_{GW}\simeq 10^{-7}$, compared to a sensitivity of $10^{-8}$.

Before concluding let us point out that our scenario provides a general source of Hierarchy that can potentially be  used to explain other Hierarchies present in late time physics, such as explaining the magnitude of  a cosmological constant that fits observations in a FLRW model\footnote{For example, consider adding to the original Lagrangian of eq.(\ref{lagrangiana}) a second potential term of the form $V_2(\phi)= M^4/f^2(\phi)$. The same mechanism that gives $M_{EW}/M_{Pl}\simeq 10^{-14}$ could also generate $\Lambda_{today}/M\simeq 10^{-14}$. It is quite remarkable that without introducing another mass scale we can generate even a tiny cosmological constant, but of course this does not solve the ``quantum'' cosmological constant problem.} or the hierarchy in the particle spectrum.

This feature can also be exploited in order to understand better what is the scale $M_{EW}$, in relation to the Little Hierarchy problem.
Usually one argues that $M_{EW}$ has to be at least $10$ TeV in order to forbid dangerous higher dimensional operators, but then one has to understand why the Higgs mass is still quite lower than $10$ TeV. A solution can be that in fact $M_{EW}$ is of the order of the Higgs mass and some symmetries forbid the dangerous operators, such as the ones which violate baryon number\footnote{It is not required to have such baryon number violating operators in order to obtain baryogenesis, see for instance \cite{Balaji}.}. 
Another solution could be to couple the higher dimensional operators as $\frac{(\bar{\psi}\psi)^2}{\phi^2}$ instead of $\frac{(\bar{\psi}\psi)^2}{M^2}$ (where $\psi$ is some SM fermion). In this way they will be extremely suppressed. Of course one needs a justifications for having this kind of couplings, but still it is a possibility that is availiable and ready to be exploited.

\section{Conclusions}

We have presented a new way of explaining the Hierarchy between $M_{Pl}$ and $M_{EW}$ starting from a theory where there is roughly one single scale $M$, but we dynamically obtain the huge splitting between the two scales. It is assumed that the Universe starts in a false vacuum that gives rise to Inflation. The presence of a scalar field 
$\phi$ with non minimal couplings to gravity gives rise to a decelerating phase after inflation, which leads to graceful exit when the false vacuum can tunnel to a zero energy vacuum. Its final value sets the value of the Planck mass, and therefore it explains the Hierarchy problem assuming that the tunneling rate per unit volume $\Gamma_{vac}$ is small enough with respect to $M^4$.

If our scenario is actually realized in nature, this would mean that all the nonrenormalizable operators of any particle physics model will start becoming important at the TeV scale, which opens up the possibility of observing completely new physics at the LHC. 

In addition future experiments on gravitational waves based on space interferometers (LISA) will definitely be able to confirm or rule out our proposal, through the detection of a clear peak with $\Omega_{GW} h^2\simeq 10^{-7}$ at frequencies of $10^{-1}-10^{-3}$ mHz.

We have also discussed  the power spectra of scalar and tensor fluctuations from inflation. Depending on the specific realization of our idea we have made two predictions.
In one case, fixing the scalar spectral index $n_S$ using the WMAP 3-year data, the tensor-to-scalar ratio is predicted to be $P_T/P_S \simeq 0.25$, detectable in the near future~\cite{Cooray}.
In the other case we have no prediction on $P_T/P_S$ but we have the prediction $n_S\simeq 0.95$ already in good agreement with the new WMAP data~\cite{WMAP3}.

The amplitude of cosmological perturbations implies some initial hierarchy ($10^{-3}-10^{-4}$), which comes from the observed small number $\delta T/T \approx 10^{-5}$ from the CMB, and from the need to stabilize the $\phi$ field at late times, thus recovering Einstein gravity.

\section*{Acknowledgments}

We are grateful to R.H.~Brandenberger, A.~Mazumdar, G.~Moore and M.~Papucci for useful discussions and comments. The work of T.B. is supported by NSERC Grant No.\ 204540.
\vs
\appendix
\section{Appendix: Exponential inflation with $n>2$} \label{n>2}
We show in this Appendix that the stage of exponential inflation is viable also in absence of the renormalizable term $n=2$. 
For any $n>2$ keeping the lowest order terms in the kinetic and the potential terms amounts to having
\be
K(\phi)=1-\bb\left(\frac{\phi}{M} \right)^n \, , \, \bar{V}\approx \Lambda \left[1-2\bb\left(\frac{\phi}{M} \right)^n\right]  \, ,
\ee
Again, for $\phi\ll M$, the canonical scalar field variable $\Phi$ and $\phi$ almost coincide ($K\approx 1$).
In the slow-roll approximation now, to lowest order in $\phi/M$, we get now:
\be
\bar{V}(\Phi)=\Lambda\left[1-2\bb\left({\Phi\over M}\right)^n\right] \, ,
\ee
and therefore:
\be
\e\equiv{M^2\over 2}\left|{1\over V}{dV\over d\Phi}\right|^2=2\bb^2n^2\left(\frac{\Phi}{M} \right)^{2(n-1)} \, ,
\ee
\be
\eta\equiv M^2{1\over V}{d^2V\over d\Phi^2}=-2 \bb n(n-1)\left(\frac{\Phi}{M} \right)^{n-2} \label{Aepseta} \, .
\ee
Now, in the slow-roll approximation, the relation between the number of efolds ${\bar{\cal N}}$ (defined to be zero when $\phi\approx M$) and $\Phi$ is given by:
\be
\Phi \approx M \left(2 n (n-2) \beta \bar{{\cal N}} \right)^{\frac{1}{2-n}}  \label{Afienne} \, ,
\ee 
and therefore the total number of e-foldings is given by
\be
\bar{{\cal N}}_{tot}\approx {1\over  2 n(n-2)\bb \left(\frac{\Phi_0}{M} \right)^{n-2}}
\label{Aenne}  \, ,
\ee 
where $\Phi_0$ refers to the initial position of $\Phi$.
Again, when $\phi$ grows to become of the order of $M$, the exponential phase ends, and there is a transition to a power-law phase.

As for the phenomenological constraints, and using eq.(\ref{Afienne}) we get now:
\be
n_S-1\simeq 2 \eta \simeq -\frac{2}{N_{3000 h^{-1} Mpc}} \left(\frac{n-1}{n-2}\right)  \, , \label{Ans}
\ee
which leads to the prediction:
\be
0.94 \leq n_S \leq 0.96 \label{prediction}
\ee
This prediction is amazingly confirmed by WMAP 3-year data (\cite{WMAP3}), which has measured $n_s=0.951+0.015-0.019$ \footnote{This comment is added in a revised version, after the release of the WMAP 3-year data}.

The slow roll parameter $\epsilon$ turns out to be negligible and it is (for example in the case $n=4$):
\be
\epsilon\approx 10^{-7}/\beta \label{eps} \, .
\ee
Given these values for the slow-roll and using eq.(\ref{ampiezza}), we get a constraint on $\lambda$:
\be
\lambda \simeq 2 \times 10^{-4} \beta^{-1/4} \, .
\ee
In this case a reasonable choice of parameters 
is again given by eq.(\ref{reasonable}).
Finally, using $\Phi_0\sim \bar{H}_I$ and eq.(\ref{Aenne}) and the numbers in eq.(\ref{reasonable}) we get typically a huge number of efolds:
\be
\bar{{\cal N}}\approx 10^{9}  \, .
\ee


\begin{thebibliography}{99}

\bibitem{lep}
  R.~Barbieri and A.~Strumia,
  arXiv:hep-ph/0007265;
  R.~Barbieri and A.~Strumia,
  Phys.\ Lett.\ B {\bf 462} (1999) 144
  [arXiv:hep-ph/9905281].

\bibitem{ADD}
  N.~Arkani-Hamed, S.~Dimopoulos and G.~R.~Dvali,
  Phys.\ Lett.\ B {\bf 429}, 263 (1998)
  [arXiv:hep-ph/9803315].

\bibitem{RS}
  L.~Randall and R.~Sundrum,
  Phys.\ Rev.\ Lett.\  {\bf 83}, 3370 (1999)
  [arXiv:hep-ph/9905221].

\bibitem{Guth}
A.~H.~Guth,
Problems,''
Phys.\ Rev.\ D {\bf 23}, 347 (1981).

\bibitem{Coleman} S.~R.~Coleman,
  Phys.\ Rev.\ D {\bf 15}, 2929 (1977)
  [Erratum-ibid.\ D {\bf 16}, 1248 (1977)];
    C.~G.~.~Callan and S.~R.~Coleman,
  Phys.\ Rev.\ D {\bf 16}, 1762 (1977);
  S.~R.~Coleman and F.~De Luccia,
  Phys.\ Rev.\ D {\bf 21}, 3305 (1980).


\bibitem{Banks}
  T.~Banks,
  arXiv:hep-th/0211160.

\bibitem{dolgovford}
  A.~D.~Dolgov, in {\it The very early Universe}, edited by G.~W.~Gibbons, S.~W.~Hawking, and S.~T.~C.~Siklos (Cambridge University Press, Cambridge, 1983); see also  L.~H.~Ford,
  Phys.\ Rev.\ D {\bf 35}, 2339 (1987).



\bibitem{chaotic}
  A.~D.~Linde,
  ``Chaotic Inflation,''
  Phys.\ Lett.\ B {\bf 129} (1983) 177.


\bibitem{LindeBook}
  A.~D.~Linde,
  {\it ``Particle Physics and Inflationary Cosmology''}, Chur, Switzerland: Harwood (1990) 362 p. (Contemporary concepts in physics, 5), 
  arXiv:hep-th/0503203.

\bibitem{hybrid}
  A.~D.~Linde,
  Phys.\ Rev.\ D {\bf 49}, 748 (1994)
  [arXiv:astro-ph/9307002].



\bibitem{Cooray}
  A.~Cooray,
  Mod.\ Phys.\ Lett.\ A {\bf 20},  (2005)
  [arXiv:astro-ph/0503118].


\bibitem{notaridimarco}
 F.~Di Marco and A.~Notari,
  Phys.\ Rev.\ D {\bf 73}, 063514 (2006)
  [arXiv:astro-ph/0511396].


\bibitem{topological}
A.~D.~Linde and D.~A.~Linde,
  Phys.\ Rev.\ D {\bf 50}, 2456 (1994)
  [arXiv:hep-th/9402115].
  A.~Vilenkin,
  Phys.\ Rev.\ Lett.\  {\bf 72}, 3137 (1994)
  [arXiv:hep-th/9402085].
\bibitem{copeland}
  E.~J.~Copeland, A.~R.~Liddle and D.~Wands,
  Phys.\ Rev.\ D {\bf 57}, 4686 (1998)
  [arXiv:gr-qc/9711068].

\bibitem{lisa}
  K.~Danzmann,
{\it Prepared for TAMA Workshop on Gravitational Wave Detection, Saitama, Japan, 12-14 Nov 1996}; http://lisa.jpl.nasa.gov.


\bibitem{GW}
  A.~Nicolis,
  Class.\ Quant.\ Grav.\  {\bf 21}, L27 (2004)
  [arXiv:gr-qc/0303084];
M.~S.~Turner and F.~Wilczek,
Phys.\ Rev.\ Lett.\  {\bf 65}, 3080 (1990);
A.~Kosowsky, M.~S.~Turner and R.~Watkins,
Phys.\ Rev.\ Lett.\ {\bf 69}, 2026 (1992);
Phys.\ Rev.\ D {\bf 45} 4514 (1992).
A.~Kosowsky and M.~S.~Turner,
Phys.\ Rev.\ D {\bf 47}, 4372 (1993)
[arXiv:astro-ph/9211004];
M.~Kamionkowski, A.~Kosowsky and M.~S.~Turner,
Phys.\ Rev.\ D {\bf 49} 2837 (1994)
[arXiv:astro-ph/9310044];
A.~Kosowsky, A.~Mack and T.~Kahniashvili,
Phys.\ Rev.\ D {\bf 66}, 024030 (2002)
[arXiv:astro-ph/0111483];
A.~D.~Dolgov, D.~Grasso and A.~Nicolis,
Phys.\ Rev.\ D {\bf 66}, 103505 (2002)
[arXiv:astro-ph/0206461].



\bibitem{moduli}
  G.~D.~Coughlan, W.~Fischler, E.~W.~Kolb, S.~Raby and G.~G.~Ross,
   Phys.\ Lett.\ B {\bf 131}, 59 (1983);
  J.~R.~Ellis, D.~V.~Nanopoulos and M.~Quiros,
  Phys.\ Lett.\ B {\bf 174}, 176 (1986);
  B.~de Carlos, J.~A.~Casas, F.~Quevedo and E.~Roulet,
  Phys.\ Lett.\ B {\bf 318}, 447 (1993)
  [arXiv:hep-ph/9308325];
  T.~Banks, D.~B.~Kaplan and A.~E.~Nelson,
  Phys.\ Rev.\ D {\bf 49}, 779 (1994)
  [arXiv:hep-ph/9308292].





\bibitem{WMAP3}
  D.~N.~Spergel {\it et al.},
  arXiv:astro-ph/0603449.

\bibitem{Balaji}
  K.~R.~S.~Balaji, T.~Biswas, R.~H.~Brandenberger and D.~London,
  Phys.\ Rev.\ D {\bf 72}, 056005 (2005)
  [arXiv:hep-ph/0506013]; 
  Phys.\ Lett.\ B {\bf 595}, 22 (2004)
  [arXiv:hep-ph/0403014].


\bibitem{BN2}
S.~Alexander, T.~Biswas and A.~Notari, in preparation.





\end{thebibliography}
\end{document}